\def\final{0}

\documentclass[superscriptaddress, 
notitlepage, amsmath, amssymb, aps,
pra,
 10pt, letterpaper]{revtex4-1}
\usepackage{graphicx}
\usepackage{dcolumn}
\usepackage{bm}

\usepackage{fancyhdr}
\usepackage{latexsym}
\usepackage{graphicx,color}
\usepackage{epstopdf}  
\usepackage[pdfstartview=FitH]{hyperref}
\usepackage{xargs}
\usepackage{shadow,epsf,amsthm,amssymb,amsmath}

\usepackage{enumerate}

\usepackage{setspace}                           

\usepackage{lipsum}

\usepackage{tikz}	
\usetikzlibrary{backgrounds,fit,decorations.pathreplacing}  

\usepackage[colorinlistoftodos,prependcaption]{todonotes}

\newtheorem{definitionenv}{Definition}
\newtheorem{lemmaenv}[definitionenv]{Lemma}
\newtheorem{theoremenv}[definitionenv]{Theorem}
\newtheorem{corollaryenv}[definitionenv]{Corollary}
\newtheorem{propositionenv}[definitionenv]{Proposition}
\newtheorem{conjectureenv}[definitionenv]{Conjecture}

\newtheorem{exampleenv}{Example}
\newtheorem{app-lemmaenv}[section]{Lemma}

\newenvironment{definition}{\begin{definitionenv}\rm}{\end{definitionenv}}
\newenvironment{lemma}{\begin{lemmaenv}\rm}{\end{lemmaenv}}
\newenvironment{theorem}{\begin{theoremenv}\rm}{\end{theoremenv}}
\newenvironment{corollary}{\begin{corollaryenv}\rm}{\end{corollaryenv}}
\newenvironment{example}{\begin{exampleenv}\rm}{\end{exampleenv}}
\newenvironment{proposition}{\begin{propositionenv}\rm}{\end{propositionenv}}
\newenvironment{conjecture}{\begin{conjectureenv}\rm}{\end{conjectureenv}}
\newenvironment{app-lemma}{\begin{app-lemmaenv}\rm}{\end{app-lemmaenv}}

\newcommand{\bd}{\begin{definition}}
\newcommand{\ed}{\end{definition}}
\newcommand{\bl}{\begin{lemma}}
\newcommand{\el}{\end{lemma}}
\newcommand{\elp}{\hspace*{\fill} $\Box$
                 \end{lemma}}
\newcommand{\bt}{\begin{theorem}}
\newcommand{\et}{\end{theorem}}
\newcommand{\etp}{\hspace*{\fill} $\Box$
                 \end{theorem}}
\newcommand{\bc}{\begin{corollary}}
\newcommand{\ec}{\end{corollary}}
\newcommand{\ecp}{\hspace*{\fill} $\Box$
                 \end{corollary}}
\newcommand{\bcj}{\begin{conjecture}}
\newcommand{\ecj}{\end{conjecture}}

\newcommand{\be}{\begin{example}}
\newcommand{\ee}{\end{example}}
\newcommand{\eep}{\hspace*{\fill} $\Box$
                 \end{example}}
\newcommand{\bp}{\begin{proposition}}
\newcommand{\ep}{\end{proposition}}
\newcommand{\epp}{
                 \end{proposition}}

\newcommand{\bra}[1]{\langle#1|}
\newcommand{\ket}[1]{|#1\rangle}
\newcommand{\braket}[2]{\langle#1|#2\rangle}

\newcommand{\tr}[1]{\text{tr}\left(#1\right)}

\newcommand{\eeq}{ \setcounter{equation} {\value{enumi}}}

\newcommand{\cA}{\mathcal{A}}
\newcommand{\cB}{\mathcal{B}}
\newcommand{\cC}{\mathcal{C}}

\newcommand{\cE}{\mathcal{E}}

\newcommand{\cM}{\mathcal{M}}

\newcommand{\sfF}{\textsf{F}}

\newcommand{\mC}{{\mathbb C}}

\def\tr{\textnormal{tr}}
\def\beq{\begin{equation}}
\def\eeq{\end{equation}}

\def\bean{\begin{IEEEeqnarray*}{rCl}}
\def\eean{\end{IEEEeqnarray*}}

\newcommand{\I}{\mathbb{I}}

\newcommand{\trnorm}[1]{\left\|#1\right\|_{\mathrm{tr}}}

\newcommand{\ot}{\otimes}
\newcommand{\eps}{\epsilon}
\newcommand{\etal}{{\it et~al.}}
\newcommand{\Hmin}{H_{\min}}

\newcommand{\kg}{\textsf{KeyGen}}

\newcommand{\dec}{\textsf{Dec}}
\newcommand{\enc}{\textsf{Enc}}

\ifnum\final=0
\newcommand{\mynote}[2]{{\color{#1} \marginpar{\tiny #2}}}
\newcommand{\mybignote}[2]{{\color{#1} $\langle \langle$ #2$\rangle \rangle$}}
\newcommandx{\rednote}[2][1=]{\todo[linecolor=red,backgroundcolor=red!25,bordercolor=red,#1]{#2}}
\newcommandx{\bluenote}[2][1=]{\todo[linecolor=blue,backgroundcolor=blue!25,bordercolor=blue,#1]{#2}}
\newcommandx{\yellownote}[2][1=]{\todo[linecolor=yellow,backgroundcolor=yellow!25,bordercolor=yellow,#1]{#2}}
\newcommandx{\greennote}[2][1=]{\todo[inline,linecolor=olive,backgroundcolor=green!25,bordercolor=olive,#1]{#2}}

\else
\newcommand{\mynote}[2]{}
\newcommand{\mybignote}[2]{}
\newcommand{\rednote}[2][1=]{}
\newcommand{\bluenote}[2][1=]{}
\newcommand{\greennote}[2][1=]{}
\newcommand{\yellownote}[2][1=]{}

\fi

\begin{document}
\title{Quantum Encryption and Generalized Quantum  Shannon Impossibility}
\author{Ching-Yi Lai}
\email{cylai@nctu.edu.tw}
\affiliation{\footnotesize Institute of Communications Engineering, National Chiao Tung University, Hsinchu 30010, Taiwan}
\author{Kai-Min Chung}
\affiliation{\footnotesize Institute of Information Science, Academia Sinica, Taipei 11529, Taiwan}



\begin{abstract}
The famous Shannon impossibility result says that any encryption scheme with perfect secrecy requires a secret key at least as long as the message.
In this paper we provide its quantum analogue with imperfect secrecy and imperfect correctness.
We also give a systematic study of information-theoretically secure quantum encryption with two secrecy definitions.
We show that the weaker one implies the stronger but with a security loss in $d$, where $d$ is the dimension of the encrypted quantum system.
This is good enough if the target secrecy error is of $o(d^{-1})$.
\end{abstract}

\maketitle

\section{Introduction}
\label{intro}

Encryption schemes are typically considered as a computational primitive, since an information-theoretically secure (ITS) symmetric key encryption scheme can only securely encrypt messages of length at most the length of the secret key  by Shannon's impossibility result\,\cite{Shan49}.
This impossibility result has been extended to the case of imperfect secrecy and imperfect correctness (see~\cite{Dodis2012}).

We start by revisiting the quantum Shannon impossibility for quantum encryption using classical keys. Let $K$ be a discrete random variable.
An encryption scheme is a quantum operation $\Psi$ that maps a quantum state $\rho$ to $\Psi_k(\rho)$ using a classical key $k\leftarrow K$ with probability $\Pr(k)$.
We say an encryption scheme is \emph{perfect} if it maps an arbitrary quantum state to a fixed state and it is invertible.
There is a lower bound on the entropy of keys that if an encryption scheme that \emph{perfectly} encrypts $n$ qubits using a classical key  drawn from $K$, then
\[
H(K)\geq 2n,
\]
where $H$ is the Shannon entropy. 
The proof was given by Boykin and Roychowdhury for the case without any initial ancilla in encryption\,\cite{BR03}. Ambainis \emph{et al.} provided a more general proof that allows an initial fixed ancilla\,\cite{AMTW00}.
(A slightly more general version, where the initial ancilla state may depend on the key, was given in~\cite{NS06b}.
See also~\cite{DHT03,Jain12} for a different proof of the same lower bound.)
For imperfect encryption, the encrypted state can deviate from a fixed state by an amount $\epsilon$, which is called  \emph{secrecy error} (see Def.~\ref{def:strongITS2}).
It was shown by Desrosiers and Dupuis~\cite{DD10}  that for an encryption scheme with {secrecy error}~$\epsilon$,
the classical key $K$ must have length\footnote{Note that the righthand side of  Eq.~(41) in~\cite{DD10} should be $(1-\eps)^2$ and hence there should be an additional factor of 2 in front of the term $\log(1-\eps)$ in the lower bound.}
\begin{align}
\log |K|\geq 2n+2\log (1-\epsilon), \label{eq:SI2}
\end{align}
{where $|K|$ denotes the size of the alphabet of $K$.}

Herein we further consider the case of \emph{imperfect correctness}, that is, the encryption is allowed to be non-invertible.
The amount that a quantum state after encryption-decryption deviates its original is called~\emph{correctness error} (see Def.~\ref{def:correctness}),
We use the proof technique in~\cite{DD10} and generalize the statement: for an encryption scheme with {secrecy error} $\epsilon$ and \emph{correctness error} $\gamma$,
the classical key $K$ must have length
\[
\log |K|\geq 2n+\log (1-\gamma-\sqrt{2\eps}).
\]


In the security definition of quantum encryption,  it should be considered that the adversary may hold a quantum system that is entangled with the message quantum state to be encrypted.
To achieve such security~(Def.~\ref{def:strongITS2}), the quantum Shannon impossibility says that a quantum encryption scheme has to use a classical key  roughly twice as long as the quantum message.
If the sender is not entangled with the adversary, Hayden~\etal~argued that
a key of $n+\log n+2\log(1/\epsilon)$ bits  is enough to approximately encrypt $n$ qubits~\cite{HLSW04}.
(See~\cite{NK06} for a discussion of the security with respect to different norms.)
Ambainis and Smith proposed one quantum encryption scheme based on $\delta$-biased set (generalized from~\cite{DS05}) that requires  a key of  $n+ 2\log n+2\log(1/\epsilon)+O(1)$ bits,
together with two other schemes that use shorter keys~\cite{AS04}.
Later Dickinson and Nayak improved this upper bound to $n+2\log(1/\eps)+O(1)$~\cite{DN06}.
The entropic notion of this security was studied by Desrosiers~\cite{Des09}.
Herein we will proceed to study this notion of security without assuming that the message quantum state is not entangled with the adversary (Def.~\ref{def:weakITS}),
which is called \emph{weak} information-theoretic security in this paper.
We show that a weak ITS quantum encryption with secrecy error $\eps$
is also ITS but with secrecy error $d^2 \eps$,
where $d$ is the dimension of the message state to be encrypted.
As an application, we can prove the security of an encryption with respect to this weak notion
when the secrecy loss is not that important. For example,  the quantum homomorphic encryption scheme for Clifford circuits
uses a key from the permutation group of $O(n\log n)$ bits, where $n=\log d$~\cite{OTF15}.
Finally we will also prove a Shannon impossibility result for this weak notion of security.



This paper is organized as follows.
In Sec.~\ref{sec:prelim} we give some basics about quantum information processing. Then we discuss the notions of quantum encryption in Sec.~\ref{sec:ITSencryption}.
We prove the quantum Shannon impossibility results in Sec.~\ref{sec:Shannon}, followed by the discussion section.

\section{Preliminaries} \label{sec:prelim}

A quantum system will be denoted by a capital letter and its corresponding complex Hilbert space will be denoted by the corresponding calligraphic letter.
For example, the Hilbert space of a quantum system $A$ is $\cA$. We will use $d_A$ to denote the dimension of $\cA$.
Let $L(\cA)$ denote the space of linear operators on a complex Hilbert space $\cA$.
A  quantum state of system $A$ is described by a \emph{density operator} $\rho\in L(\cA)$ that is positive semidefinite  and with unit trace $(\tr(\rho)=1)$. 
Let $D(\cA)= \{ \rho\in L(\cA): \rho\geq 0, \tr(\rho)=1\}$ be the set of density operators on  $\cA$.
When $\rho\in D(\cA)$ is of rank one, it is called a \emph{pure} quantum state and we can write $\rho=\ket{\psi}\bra{\psi}$ for some unit vector $\ket{\psi}\in \cA$,
where $\bra{\psi}=\ket{\psi}^{\dag}$ is the conjugate transpose of $\ket{\psi}$. If  $\rho$ is not pure, it is called a \emph{mixed} state and can be expressed as a convex combination of pure quantum states.

Associated with an $n$-qubit quantum system is a complex Hilbert space $\mC^{2^n}$ with a computational basis $\{\ket{v}:v\in \{0,1\}^n\}$.
Pauli matrices $\sigma_0=\begin{bmatrix}1 &0\\0&1\end{bmatrix}$, $\sigma_1=\begin{bmatrix}0 &1\\1&0\end{bmatrix}$,  $\sigma_3=\begin{bmatrix}1 &0\\0&-1\end{bmatrix}$, and $\sigma_2=\begin{bmatrix}0 &-i\\i&0\end{bmatrix}$ are a basis for linear operators on a single qubit.
Consequently $n$-fold Pauli matrices form a basis for linear operators on $n$ qubits.

When $\rho\in D(\cA)$ is diagonal in the computational basis $\{\ket{0}, \ket{1}, \dots, \ket{d_A-1}\}$,
we say $\rho$ is classical. In this case, $\rho$ corresponds to a discrete distribution on $\{0,1,\dots,d_A-1\}$.
In other words, a random variable of $n$ classical bits can be represented as an $n$-qubit quantum state that is diagonal in the computational basis.

The Hilbert space of a joint quantum system $AB$ is the tensor product of the corresponding Hilbert spaces $\cA\otimes \cB$.
We will use a subscript to specify which subsystem a vector belongs to or an operator operates on.
Let $\I_A$ denote the identity on the subsystem $A$.
Thus the totally-mixed state in $A$ is $\frac{1}{d_A}\I_A$.
For $\rho_{AB}\in D(\cA\ot \cB)$, we will use  $\rho_A=\tr_B(\rho_{AB})$
to denote its reduced density operator in system $A$,
where the \emph{partial trace} is defined by
\[
\tr_B(\rho_{AB})= \sum_{i} \left(\I_A\ot\bra{i}_B\right) \rho_{AB}\left( \I_A\ot \ket{i}_B\right)
\]
for an orthonormal basis $\{\ket{i}_B\}$ for $\cB$.

Suppose $\rho_A\in D(\cA)$ of finite dimension $d_A$. Then there exists $\cB$ of dimension $d_B\geq d_A$ and $\ket{\psi}_{AB}\in \cA\otimes \cB$ such that
\[
\tr_B \ket{\psi}_{AB}\bra{\psi} = \rho_A.
\]
Such $\ket{\psi}_{AB}$ is called a \emph{purification} of $\rho_A$~\cite{HJW93}.

A \emph{separable} state $\rho_{AB}$ has a density operator of the form
\[
\rho_{AB}=\sum_{x} p_x \rho_A^x\otimes \rho_B^x,
\]
where $\rho_A^x \in D(\cA)$ and $\rho_B^x \in D(\cB)$.
Especially when $A$ is classical, \[
\rho_{AB}=\sum_{x} p_x \ket{x}_A\bra{x}\otimes \rho_B^x,
\]
is called a classical-quantum (cq) state.

The evolution of a quantum state $\rho\in D(\cA)$ is described by a completely positive and trace-preserving  map $\Psi: D(\cA)\rightarrow D(\cA')$ with operation elements $\{E_j\}$:
\begin{align}\Psi(\rho)=\sum_j E_j \rho E_j^\dag, \ \sum_j E_j^\dag E_j=\I_A.\label{eq:QO}
\end{align}
{In particular, if the evolution is a unitary $U$, we have the evolved state $U(\rho)=U\rho U^{\dag}$.
}

The trace distance between two quantum states $\rho$ and $\sigma$ is
$$||{\rho}-{\sigma}||_{\mathrm{tr}},$$
where $||X||_{\mathrm{tr}}=\frac{1}{2}\tr{\sqrt{X^{\dag}X}}$ is the trace norm of $X$.
The fidelity between $\rho$ and $\sigma$ is
$$
F(\rho,\sigma)=\tr \sqrt{\rho^{1/2}{\sigma}\rho^{1/2}}.
$$
If $\rho=\ket{\psi}\bra{\psi}$,
\begin{align}
F(\ket{\psi}\bra{\psi},\sigma)=\sqrt{\bra{\psi}\rho\ket{\psi}}.\label{eq:pureF}
\end{align}
\bt[Uhlmann's theorem~\cite{Uhl76}]
$$F(\rho_{A},\sigma_A)= \max_{\ket{\psi},\ket{\phi}} |\braket{\psi}{\phi}|= \max_{\ket{\phi}} |\braket{\psi'}{\phi}|,$$ where the maximization is over all purifications $\ket{\phi}$ of $\sigma_A$ and all purifications $\ket{\psi}$ of $\rho_A$ or any fixed purification $\ket{\psi'}$ of $\rho_A$.
\et
\noindent Below is a variant of Uhlmann's theorem.
\bc \label{cor:Uhlmann}
Suppose $\rho_A$ is a reduced density operator of $\rho_{AB}$.
Suppose $\rho_A$ and $\sigma_A$ have fidelity $F(\rho_A,\sigma_A)\geq 1- \epsilon$. Then
there exists $\sigma_{AB}$ with $\tr_B(\sigma_{AB})=\sigma_A$ such that $F(\rho_{AB},\sigma_{AB})\geq 1- \epsilon$.
\ec
\begin{proof}
Let $\ket{\psi}_{ABR}$ be a purification of $\rho_{AB}$, which is immediately  a purification of $\rho_A$.
Since $F(\rho_A,\sigma_A)\geq 1- \epsilon$, suppose $\ket{\phi}_{ABR}$ is a purification of $\sigma_A$ such that $|\braket{\psi}{\phi}|\geq 1-\epsilon$. Let $\sigma_{AB}= \tr_R(\ket{\phi}\bra{\phi})$.
Then $F(\rho_{AB},\sigma_{AB})\geq |\braket{\psi}{\phi}| \geq 1- \epsilon$.
\end{proof}

A relation between the fidelity and the trace distance of two quantum states was proved by Fuchs and van~de Graaf~\cite{FvdG99} that
\begin{align}
1-F(\rho,\sigma)\leq   || \rho-\sigma||_{\mathrm{tr}}\leq \sqrt{1- F^2(\rho,\sigma)}. \label{eq:FT}
\end{align}
When $\rho$ or $\sigma$ is pure, we have
\begin{align}
  || \rho-\sigma||_{\mathrm{tr}}\geq {1- F^2(\rho,\sigma)}. \label{eq:pureFT}
\end{align}

The   min-entropy of $A$ conditioned on $B$ is defined as
$$
    \Hmin(A|B)_{\rho} =  -\inf_{\sigma_{B}}
    \left\{
      \inf \left\{\lambda\in \mathbb{R}: \rho_{AB}\leq 2^\lambda \I_{A}\otimes \sigma_B\right\}
    \right\}.
 $$
K\"{o}nig \emph{et al}. gave an operational definition of min-entropy~\cite{KRS09}:
\begin{align}
    2^{-H_{\min}(A|B)_{\rho}} =  d_A \max_{\Psi} F\left( ( \I_A\otimes \Psi_B) (\rho_{AB}), \ket{\Phi}_{AB'}\bra{\Phi}_{AB'}\right),\label{eq:Hmin}
\end{align}
 where $\Psi: B\rightarrow B'$ is a quantum operation as defined in Eq.~(\ref{eq:QO}), $B'\equiv A$, and $\ket{\Phi}_{AB'}= \frac{1}{\sqrt{d_A}}\sum_{x}\ket{x}_A\ket{x}_{B'}$ is the maximally-entangled state.
The leakage chain rule for conditional min-entropy when $\rho_{AB}$ is separable~\cite{DD10} says that
\begin{align}
H_{\min}(A|B)_{\rho}\geq H_{\min}(A)- \log d_B. \label{eq:leakage}
\end{align}

\section{Information-theoretically Secure Quantum Encryption}\label{sec:ITSencryption}

Suppose the sender holds a quantum system $M$, which  may or may not be entangled with a quantum system $E$ held by the adversary.
The joint system of the sender and the adversary will be denoted by the subscript $\rho_{ME}$.

\begin{definition} \label{def:encryption}
A symmetric-key quantum encryption  scheme $\sfF$ is defined by the following   algorithms:\\
1)   (Key generation) $\sfF.\kg$: $1^{\eta}\rightarrow  K $, where $\eta$ is a security parameter and $K$ is a discrete random variable $\{k, \Pr(k)\}$ depending on $\eta$. The algorithm  outputs a \emph{classical} private key $k\in K$.

\noindent 2)  (Encryption) $\sfF.\enc_{k}$: $D(\cM)\rightarrow D(\cC)$. 
The algorithm takes a secret key $k$ and a quantum state   $\rho\in D(\cM)$ as input and outputs a ciphertext  $\sigma \in D(\cC)$.

\noindent 3) (Decryption) $\sfF.\dec_{k}$: $D(\cC)\rightarrow D(\cM)$.
   The algorithm takes $k$ and $\sigma\in D(\cC)$ as input and outputs a  quantum state $\hat{\rho}\in D(\cM)$.
\end{definition}

We will use $\Gamma_{\sfF.\text{func}}^{\textsf{arg}}$ to denote the quantum operation corresponding to the algorithm $\sfF.\text{func}$ with  argument $\textsf{arg}$.
For example, $\Gamma_{\sfF.\enc}^{k}$ is the encryption algorithm of $\sfF$ using a secret key $k$ generated from the distribution $K$.
In addition, when $\textsf{arg}$ is a distribution, it means the quantum operation is averaged over the distribution. Consequently,
 $\Gamma_{\sfF.\enc}^{K}$ denotes the encryption algorithm of $\sfF$ weighted by the probability distribution $K$, that is,   \[
\Gamma_{\sfF.\enc}^{K}(\rho_M) =  \sum_{k\in K} \Pr(k) \Gamma_{\sfF.\enc}^{k}(\rho_M).
\]
The encrypted quantum state in the view of the adversary is then
\[
 \Gamma_{\sfF.\enc}^{K} \otimes \mathbb{I}_E (\rho_{ME}).
\]

\begin{definition} \label{def:correctness}
\noindent (Correctness)
 $\sfF$ is an encryption scheme with \emph{correctness error} $\gamma=\gamma(\eta)$ if    for any $\rho_{ME}\in D(\cM\otimes \cE )$
and $k\leftarrow \sfF.\kg(1^{\eta})$,
\begin{align}
&\left\| { \Gamma_{\sfF.\dec}^{k} \otimes \mathbb{I}_E \left(   \Gamma_{\sfF.\enc}^{k}\otimes \mathbb{I}_E(\rho_{ME})\right)} -  \rho_{ME} \right\|_{\mathrm{tr}} \leq  \gamma. \label{eq:1}
\end{align}
\end{definition}

Semantic security in the computational setting was introduced by Goldwasser and Micali~\cite{GM84}.
Following that, an information-theoretic security notion  of entropy security was defined (in a classical setting) by Russel and Wang\,\cite{RW06}.
Dodis and Smith\,\cite{DS05} discussed the security notions more generally  and showed that entropy security is equivalent to a security notion of indistinguishability.
Several ITS notions of perfect security are further discussed in~\cite{IO11,IOS18}.
Entropic security and entropic indistinguishability are also equivalent~in the quantum settings~\cite{Des09,DD10}.
It is natural to use trace distance as a measure of the indistinguishability of two quantum states.

\begin{definition}\label{def:strongITS2}
\noindent (Secrecy)
 $\sfF$ is an ITS encryption scheme with \emph{secrecy error} $\epsilon=\epsilon(\eta)$ if  there exists $\Omega_M\in D(\cM)$ such that for any $\rho_{ME}\in D(\cM\otimes \cE)$,
  \begin{align}
 \left\|     \Gamma_{\sfF.\enc}^{K} \otimes \mathbb{I}_E (\rho_{ME}) -   \Omega_M \otimes \rho_{E} \right\|_{\mathrm{tr}}
&\leq  \epsilon.  \label{eq:2.2}
\end{align}
\end{definition}

This security definition says that an encrypted quantum state of an ITS encryption is statistically indistinguishable
to the tensor product of a fixed quantum state and the local state of the adversary.

We can also define the security as follows.
\begin{definition}\label{def:strongITS}
\noindent (Secrecy)
 $\sfF$ is an ITS encryption scheme with \emph{secrecy error} $\epsilon=\epsilon(\eta)$ if    for any $\rho_{ME}, \rho'_{ME}\in D(\cM\otimes \cE)$ with $\tr_M (\rho_{ME})=\tr_{M}(\rho'_{ME})$,
  \begin{align}
 \left\|     \Gamma_{\sfF.\enc}^{K} \otimes \mathbb{I}_E (\rho_{ME}) -    \Gamma_{\sfF.\enc}^{K} \otimes \mathbb{I}_E(\rho'_{ME}) \right\|_{\mathrm{tr}}
&\leq  \epsilon.  \label{eq:2}
\end{align}
\end{definition}
Definition~\ref{def:strongITS2} directly implies Def.~\ref{def:strongITS} with secrecy error $2\epsilon$ by the triangle inequality.
On the other hand, consider $\rho_{ME}'=\ket{0}_E\bra{0}\otimes \rho_E$, where $\rho_E= \tr_M \rho_{ME}$.
Then Def.~\ref{def:strongITS} directly implies Def.~\ref{def:strongITS2}  with $\Omega_M=   \Gamma_{\sfF.\enc}^{K}\left(\ket{0}_M\bra{0}\right)$ and secrecy error $\eps$.
Therefore, these two definitions are equivalent up to a factor of $2$.
As a consequence, we will use Def.~\ref{def:strongITS} for secrecy error
and operationally we use the one that is  more convenient for the context.

We can also define a weaker security without any reference involved.
\begin{definition}(Weak Secrecy) \label{def:weakITS}
 $\sfF$ is a \emph{weak} ITS encryption scheme with secrecy error $\epsilon=\epsilon(\eta)$ if    for $\rho, \rho'\in D(\cM)$,
  \begin{align}
 \left\|     \Gamma_{\sfF.\enc}^{K} (\rho) -    \Gamma_{\sfF.\enc}^{K}(\rho') \right\|_{\mathrm{tr}}
&\leq  \epsilon.  \label{eq:2}
\end{align}
\end{definition}
For a perfect encryption scheme, the two notions are equivalent.
If we assume that the message state is not entangled with the adversary, then the weak secrecy is equivalent to the regular secrecy.
It is easy to show that if $\rho_{ME}$ is separable, the two notions are equivalent.
\textbf{Remark}
As in the standard language of cryptography, we define the secret key $K$, correctness error $\gamma$, and secrecy error $\eps$
as functions of the security parameter $\eta$. In the following discussion we will omit the security parameter as it is irelevent to our discussion.

Next we show that an weak ITS encryption scheme is also ITS but with
an additional security loss in the dimension of the message system to be encrypted.
Before we prove this statement, we need the following lemma.
\bl \label{lemma:offdiagonal}
If $\sfF$ is a weak ITS encryption scheme with security error at most $\epsilon(\kappa)$, then, for $\braket{x}{y}=0$, we have the following mathematical result:
\[
\left\|   \Gamma_{\sfF.\enc}^{K} (\ket{x}\bra{y})\right\|_{\mathrm{tr}}\leq \epsilon(\kappa).
\]
\el
\begin{proof}
Assume $\braket{x}{y}=0$.
Consider $\ket{\psi_0}= \frac{1}{\sqrt{2}}(\ket{x}+\ket{y})$, $\ket{\psi_1}= \frac{1}{\sqrt{2}}(\ket{x}-\ket{y})$,
 $\ket{\psi_2}= \frac{1}{\sqrt{2}}(\ket{x}+i\ket{y})$,
 and $\ket{\psi_3}= \frac{1}{\sqrt{2}}(\ket{x}-i\ket{y})$. Since $\sfF$ is weak ITS,
\begin{align*}
\epsilon(\kappa)\geq& \left\|   \Gamma_{\sfF.\enc}^{K} (\ket{\psi_0}\bra{\psi_0})-   \Gamma_{\sfF.\enc}^{K} (\ket{\psi_1}\bra{\psi_1})\right\|_{\mathrm{tr}}
=\left\|   \Gamma_{\sfF.\enc}^{K} (\ket{x}\bra{y})+   \Gamma_{\sfF.\enc}^{K} (\ket{y}\bra{x})\right\|_{\mathrm{tr}},\\
\epsilon(\kappa)\geq& \left\|   \Gamma_{\sfF.\enc}^{K} (\ket{\psi_2}\bra{\psi_2})-   \Gamma_{\sfF.\enc}^{K} (\ket{\psi_3}\bra{\psi_3})\right\|_{\mathrm{tr}}
=\left\|   \Gamma_{\sfF.\enc}^{K} (i\ket{y}\bra{x})-   \Gamma_{\sfF.\enc}^{K} (i\ket{x}\bra{y})\right\|_{\mathrm{tr}}.
\end{align*}
Thus
\begin{align*}
&\left\|   \Gamma_{\sfF.\enc}^{K} (\ket{x}\bra{y})\right\|_{\mathrm{tr}}\\
=&\left\| \frac{1}{2}  \Gamma_{\sfF.\enc}^{K} (\ket{x}\bra{y})+\frac{1}{2}  \Gamma_{\sfF.\enc}^{K} (\ket{y}\bra{x})\right.
\left.+\frac{1}{2}  \Gamma_{\sfF.\enc}^{K} (\ket{x}\bra{y})-\frac{1}{2}  \Gamma_{\sfF.\enc}^{K} (\ket{y}\bra{x})\right\|_{\mathrm{tr}}\\
\leq&\frac{1}{2}\left\|   \Gamma_{\sfF.\enc}^{K} (\ket{x}\bra{y})+   \Gamma_{\sfF.\enc}^{K} (\ket{y}\bra{x})\right\|_{\mathrm{tr}}
+ \frac{1}{2}\left\|   \Gamma_{\sfF.\enc}^{K} (\ket{x}\bra{y})-   \Gamma_{\sfF.\enc}^{K} (\ket{y}\bra{x})\right\|_{\mathrm{tr}}\\
\leq& \epsilon(\kappa).
\end{align*}
\hspace*{\fill} $\Box$
\end{proof}

\bt \label{thm:weaktostrong}
If $\sfF$ is a weak ITS encryption scheme with security error at most~$\epsilon $, then $\sfF$ is a regular ITS encryption scheme with   secrecy error at most $d^2 \epsilon $,
where $d$ is the dimension of the encrypted message system.
\et
\begin{proof}

Suppose that  $\sfF$ is a weak ITS encryption scheme by Def.~\ref{def:weakITS} with secrecy error $\epsilon$.
We first consider a pure state $\rho_{ME}=\ket{\psi}_{ME}\bra{\psi}$ with $\ket{\psi}_{ME}=\sum_{i} \sqrt{\lambda_i} \ket{i}_M \ket{i}_E$
and $\rho_E=\tr_M(\ket{\psi}_{ME}\bra{\psi})=\sum_{i}\lambda_i \ket{i}_E\bra{i}$.
We would like to show that
\begin{align}
\left\|     \Gamma_{\sfF.\enc}^{K} \otimes \mathbb{I}_E (\rho_{ME}) -    \Gamma_{\sfF.\enc}^{K} (\ket{0}_M\bra{0})\otimes\rho_E \right\|_{\mathrm{tr}}\leq d^2 \epsilon. \label{eq:4}
\end{align}
If this equation holds for any pure $\rho_{ME}$, then for $\sigma_{ME}= \sum_j p_j \rho_{ME}^j$, where $\rho_{ME}^j$ are pure,
we have
\begin{align*}
&\left\|     \Gamma_{\sfF.\enc}^{K} \otimes \mathbb{I}_E (\sigma_{ME}) -    \Gamma_{\sfF.\enc}^{K} (\ket{0}_M\bra{0})\otimes \left(\sum_j p_j \rho_E^j\right) \right\|_{\mathrm{tr}}\\
\leq & \sum_j p_j  \left\|     \Gamma_{\sfF.\enc}^{K} \otimes \mathbb{I}_E (\rho_{ME}^j) -    \Gamma_{\sfF.\enc}^{K} (\ket{0}_M\bra{0})\otimes\rho_E^j \right\|_{\mathrm{tr}}\\
\leq &d^2 \epsilon.
\end{align*}
It remains to prove Eq.~(\ref{eq:4}).
  \begin{align*}
&\left\|     \Gamma_{\sfF.\enc}^{K} \otimes \mathbb{I}_E (\rho_{ME}) -    \Gamma_{\sfF.\enc}^{K}(\ket{0}_M\bra{0}) \otimes \rho_{E}) \right\|_{\mathrm{tr}}\\
=&\left\|   \sum_{i,j} \sqrt{\lambda_i\lambda_j}  \Gamma_{\sfF.\enc}^{K} (\ket{i}_M\bra{j})\otimes \ket{i}_E\bra{j} -  \Gamma_{\sfF.\enc}^{K} (\ket{0}_M\bra{0})\otimes \sum_i \lambda_i \ket{i}_E\bra{i} \right\|\\
\stackrel{}{\leq}&  \sum_i  \lambda_i \left\|   \Gamma_{\sfF.\enc}^{K} (\ket{i}_M\bra{i}) \otimes \ket{i}\bra{i} -  \Gamma_{\sfF.\enc}^{K}  (\ket{0}\bra{0})\otimes  \ket{i}_M\bra{i} \right\|_{\mathrm{tr}}\\
& +  \sum_{i\neq j}  \left\| \Gamma_{\sfF.\enc}^{K} (\ket{i}_M\bra{j})\otimes \ket{i}_E\bra{j}\right\|_{\mathrm{tr}}\\
\stackrel{(a)}{\leq}&  \sum_i  \lambda_i \left\|   \Gamma_{\sfF.\enc}^{K} (\ket{i}_M\bra{i})  -  \Gamma_{\sfF.\enc}^{K}  (\ket{0}\bra{0}) \right\|_{\mathrm{tr}}
+  \sum_{i\neq j}  \left\| \Gamma_{\sfF.\enc}^{K} (\ket{i}\bra{j})\right\|_{\mathrm{tr}}\\
\stackrel{(b)}{\leq}&d^{2}\epsilon,
\end{align*}
where $(a)$ is because $\trnorm{A\ot \ket{i}\bra{j}}=\frac{1}{2}\sqrt{A^\dag A\otimes \ket{j}\bra{j}}=\trnorm{A}$ and $(b)$ is by Def.~\ref{def:weakITS} and Lemma~\ref{lemma:offdiagonal}. \hspace*{\fill} $\Box$
\end{proof}


As an application of this theorem, it suffices to  prove that an encryption scheme is ITS with respect to Def.~\ref{def:strongITS}
if we target at secrecy error $o(2^{-n})$, where $n$ is the number of encrypted qubits.
\be
Ambainis and Smith introduced an approximate quantum encryption scheme  based on \emph{$\delta$-based sets}~\cite{AS04}.
It was shown that the Ambainis-Smith scheme is ITS with secrecy error $\delta 2^{n}$~\cite{DD10}
On the other hand, it is  weak ITS with secrecy error $\delta 2^{0.5n}$~\cite{AS04}
and hence ITS with secrecy error at most $\delta 2^{2.5 n}$ by Theorem~\ref{thm:weaktostrong}.
\ee
\be
The secrecy of the quantum homomorphic encryption scheme for Clifford circuits by Ouyang \emph{et al.}~\cite{OTF15} was proved with respect to the weak secrecy   (Def.~\ref{def:weakITS})
but we can prove its IT-security with respect to Def.~\ref{def:strongITS} without much loss in parameters by Theorem~\ref{thm:weaktostrong}.
Similarly for the secrecy of  the quantum homomorphic encryption scheme  for the instantaneous quantum polynomial (IQP) circuits in~\cite{LC18}.

\ee

At the end of this section we  prove a theorem, which is similar to \cite[Theorem 5.2]{AMTW00}, saying that an $n$-qubit imperfect encryption scheme can encrypt $2n$ classical bits
with the same correctness and secrecy errors.
\bt \label{lemma:qtoc}
 Suppose $\sfF=(\kg,\enc,\dec)$ is an ITS quantum encryption scheme on $\cM=\mathbb{C}^{2^n}$ with correctness error $\gamma$ and secrecy error $\epsilon$.
 Then  there exists an ITS quantum encryption scheme $\sfF'$ on $\cM'=\{0,1,2,3\}^{n}$ with correctness error $\gamma$ and secrecy error $\epsilon$.
\et
\begin{proof}
Suppose $\sfF$ is ITS by Def~\ref{def:strongITS}.
Define $U\ket{x}= \sigma^x \otimes \mathbb{I} \left(\frac{1}{\sqrt{2^{n}}}\sum_{i\in\{0,1\}^n}\ket{i}_A\ket{i}_B\right)$ for $x\in\{0,1,2,3\}^n$,
where $\sigma^x= \bigotimes_{j=1}^n\sigma_{x_j}$.
That is, $U$ maps $x$ to one of the $2n$-qubit Bell states.

Let $\sfF'.\enc^k: \{0,1,2,3\}^{n}\rightarrow D(\mathbb{C}^{2^{2n}})$ be the algorithm corresponding to the following operation:
 $$\Gamma_{\sfF'.\enc}^{k}=(\Gamma_{\sfF.\enc}^{k}\otimes \mathbb{I})\circ U.$$

Let $\tilde{U}$ denote the inverse of $U$, which does the Bell measurement such that $\tilde{U}( U\ket{x}\bra{x}U^\dag) =\ket{x}\bra{x}$.
Note that $U$ can be implemented by a unitary circuit, and hence $\tilde{U}$ can be implemented by the reverse of that unitary circuit.
Let $\sfF'.\dec^k:  D(\mathbb{C}^{2^{2n}})\rightarrow \{0,1,2,3\}^n$ be the algorithm corresponding to the following operation:
 $$\Gamma_{\sfF'.\dec}^{k}=\tilde{U}\circ(\Gamma_{\sfF.\dec}^{k}\otimes \mathbb{I}).$$
When the message  is classical,  the joint state of the sender and the adversary $\rho_{ME}$ is a cq state.
Thus the correctness of the encryption on the message alone implies the general correctness in Def.~\ref{def:correctness}.
Therefore,
\begin{align*}
 &\left\|     \Gamma_{\sfF'.\dec}^{k}\circ \Gamma_{\sfF'.\enc}^{k}(\ket{x}\bra{x}) -  \ket{x}\bra{x} \right\|_{\mathrm{tr}}\\
=& \left\|    \tilde{U}\circ \left({\Gamma_{\sfF.\dec}^{k} } \otimes \mathbb{I}_B \right)\circ \left(  {\Gamma_{\sfF.\enc}^{k} } \otimes \mathbb{I}_B \right)(U\ket{x}\bra{x}U^\dag) - \ket{x}\bra{x} \right\|_{\mathrm{tr}}\\
=& \left\|    \left({\Gamma_{\sfF.\dec}^{k} } \otimes \mathbb{I}_B \right)\circ \left(  {\Gamma_{\sfF.\enc}^{k} } \otimes \mathbb{I}_B \right)(U\ket{x}\bra{x}U^\dag) - U\ket{x}\bra{x}U^\dag \right\|_{\mathrm{tr}}\\
\leq&  \gamma,
\end{align*}
by the correctness of $F$.

{Similarly, for the encryption of classical messages, the weak security in~Def.~\ref{def:weakITS} implies~the regular security in Def.~\ref{def:strongITS}.}
For $v,v'\in\{0,1,2,3\}^n$,
\begin{align*}
 \left\|     \Gamma_{\sfF'.\enc}^{K}(\ket{v}\bra{v}) -    \Gamma_{\sfF'.\enc}^{K} (\ket{v'}\bra{v'}) \right\|_{\mathrm{tr}}
=& \left\|     {\Gamma_{\sfF.\enc}^{K} } \otimes \mathbb{I}_B (\rho_{AB}) -    {\Gamma_{\sfF.\enc}^{K} } \otimes \mathbb{I}_B(\rho'_{AB}) \right\|_{\mathrm{tr}}\\
\leq &  \epsilon,
\end{align*}
where $\rho_{AB}=\frac{1}{2^n}\sum_{i,j\in\{0,1\}^n} \sigma^{v}\ket{i}\bra{j}\left(\sigma^{v}\right)^{\dag}\otimes \ket{i}\bra{j}$ and similarly for $\rho'_{AB}$.
Note that $\tr_A \rho_{AB}=\tr_A \rho_{AB}'$.
The last inequality follows from the secrecy of $\sfF$ (Def~\ref{def:strongITS}).
\hspace*{\fill} $\Box$
\end{proof}


\section{Quantum Shannon Impossibility} \label{sec:Shannon}
In this section we  generalize Shannon's impossibility in the quantum case.

\bt \label{thm:shannon} Suppose $\sfF=(\kg,\enc,\dec)$ is an ITS quantum encryption scheme on $\cM=\mathbb{C}^{2^n}$ with correctness error $\gamma$ and secrecy error $\epsilon$.
Then
\[
\log |K|\geq  2n+ \log(1-\gamma-\sqrt{2\eps}).
\]

\et
\begin{proof}

Let $$\rho_{MC}=\I_M\ot \Gamma^K_{\sfF.\enc}\left( \ket{\Phi}_{MC}\bra{\Phi}\right)= \frac{1}{2^{n}|K|}\sum_{m,m'\in\{0,1\}^{n}\atop k\in K} \ket{m}_M\bra{m'}\otimes \Gamma^k_{\sfF.\enc}(\ket{m}_C\bra{m'}),$$
where $\ket{\Phi}_{MC}$ is the maximally-entangled state.
Let
 $$\sigma_{MC}=\frac{1}{2^n}\I_M\ot \Gamma^K_{\sfF.\enc}\left( \ket{0}_C\bra{0}\right)=\frac{1}{2^{n}|K|}\sum_{m\in\{0,1\}^{n}\atop k\in K} \ket{m}\bra{m}\otimes \Gamma^k_{\sfF.\enc}(\ket{0}_C\bra{0}).$$
By the secrecy of $\sfF$, we have
$
\left\| \rho_{MC}-\sigma_{MC} \right\|_{\mathrm{tr}}\leq \epsilon
$
and hence by Eq.~(\ref{eq:FT}),
\begin{align}
F(\rho_{MC}, \sigma_{MC})\geq 1-\epsilon. \label{eq:3}
\end{align}
Let $$\rho_{MCK}=\frac{1}{2^{2n}}\sum_{m,m'\in\{0,1\}^{2n}\atop k\in K}p_k \ket{m}\bra{m'}\otimes \Gamma^k_{\sfF.\enc}(\ket{m}\bra{m'})\otimes \ket{k}\bra{k},$$
which satisfies $\tr_K \rho_{MCK}=\rho_{MC}.$
By Uhlmann's theorem (Corollary~\ref{cor:Uhlmann}) and Eq.~(\ref{eq:3}), there exists $\sigma_{MCK}$  with $\tr_{K}(\sigma_{MCK})=\sigma_{MC}$ such that
\begin{align*}
F(\rho_{MCK}, \sigma_{MCK})\geq 1-\epsilon.
\end{align*}
Let $V=\sum_k \ket{kk}_{KK'}\bra{k}_K$
and let $\tau_{MCK}=\tr_{K'} V \sigma_{MCK} V^\dag$.
This ensures that $\tau_{MCK}$ is classical on $K$ and it still holds that $\tr_K\tau_{MCK}=\sigma_{MC}$.
  Then we have
\begin{align*}
1-\epsilon\leq& F(\rho_{MCK}, \sigma_{MCK})\\
=&  F(V\rho_{MCK}V^{\dag}, V\sigma_{MCK}V^\dag)\\
\leq&  F(\rho_{MCK}, \tau_{MCK})\\
\leq&  F(\I_M\ot \Gamma_{F.\dec,K}\left(\rho_{MCK}\right), \I_M\ot\Gamma_{F.\dec,K}\left(\tau_{MCK}\right))\\
\leq&  F(\tr_K \I_M\ot \Gamma_{F.\dec,K}\left(\rho_{MCK}\right),\tr_K \I_M\ot\Gamma_{F.\dec,K}\left(\tau_{MCK}\right)).
\end{align*}
where $ \Gamma_{F.\dec,K}$ is the decryption operation controlled by  the $K$ subsystem.
Hence $\left\|\tr_K \I_M\ot\Gamma_{F.\dec,K}\left(\rho_{MCK}\right)- \tr_K \I_M\ot\Gamma_{F.\dec,K}\left(\tau_{MCK}\right)\right\|_{\mathrm{tr}}\leq \sqrt{2\epsilon}.$
Thus
\begin{align*}
&\left\| \ket{\Phi}_{MC}\bra{\Phi}- \tr_K \I_M\ot\Gamma_{F.\dec,K}\left(\tau_{MCK}\right)\right\|_{\mathrm{tr}}\\
\leq & \left\| \ket{\Phi}_{MC}\bra{\Phi}- \tr_K \I_M\ot\Gamma_{F.\dec,K}\left(\rho_{MCK}\right)\right\|_{\mathrm{tr}}\\
&+ \left\|  \tr_K \I_M\ot\Gamma_{F.\dec,K}\left(\rho_{MCK}\right)-  \tr_K \I_M\Gamma_{F.\dec,K}\left(\tau_{MCK}\right)\right\|_{\mathrm{tr}}\\
\leq& \gamma +\sqrt{2\epsilon},
\end{align*}
which implies, by Eq.~(\ref{eq:pureFT}),
\begin{align*}
 1-(\gamma +\sqrt{2\epsilon})\leq& F\left( \ket{\Phi}_{MC}\bra{\Phi} , \tr_K\I_M\ot \Gamma_{F.\dec,K}\left(\tau_{MCK}\right)\right)^2\\
\stackrel{(a)}{\leq} & \frac{1}{2^n}2^{-H_{\min}(M|CK)_{\tau}}\\
\stackrel{(b)}{\leq} &2^{-n}(2^{-n+\log |K|} ),
\end{align*}
where $(a)$ is by Eq.\,(\ref{eq:Hmin}); $(b)$ follows from the leakage chain rule Eq.\,(\ref{eq:leakage}) and that $\tr_K\tau_{MCK}=\sigma_{MC}$ and $H_{\min}(M|C)_\tau=n$.
Therefore, we have
\[
\log|K|\geq 2n+ \log\left(1-\left(\gamma+\sqrt{2\epsilon}\right)\right).
\]\hspace*{\fill} $\Box$
\end{proof}

Next we consider a quantum Shannon impossibility for quantum encryption with weak IT-security in Def.~\ref{def:weakITS}.
\bt \label{thm:weakShannon_min}
Suppose $\sfF=(\kg,\enc,\dec)$ is a weak ITS quantum encryption scheme with correctness error $\gamma$ and secrecy error $\epsilon=o(2^{-n})$.
  Then the length of the classical key $K$ satisfies
  \begin{align*}
\log|S|\geq 2n+ \log\left(1-\left(\gamma+\sqrt{2^{n}\epsilon}\right)\right).
  \end{align*}
\et
\begin{proof}
Let $\rho_{AB}=\frac{1}{2^n}\I_A\otimes \Gamma_{\sfF.\enc}^{K} (\ket{\Phi}_{AB}\bra{\Phi})$ and $\sigma_{AB}=\frac{1}{2^n}\I_A\otimes \Gamma_{\sfF.\enc}^{K} (\ket{0}_B\bra{0})$.
Then \begin{align*}
\left\|\rho_{AB}- \sigma_{AB}\right\|_{\mathrm{tr}}
\leq&\frac{1}{2^n}\left\|\sum_{m,l} \ket{m}\bra{l}\otimes \Gamma_{\sfF.\enc}^{K} (\ket{m}\bra{l})\right.
\left.- \sum_m \ket{m}\bra{m}\otimes \Gamma_{\sfF.\enc}^{K} (\ket{0}\bra{0})\right\|_{\mathrm{tr}}\\
\leq&\frac{1}{2^n}\sum_{m\neq l} \left\| \Gamma_{\sfF.\enc}^{K} (\ket{m}\bra{l})\right\|_{\mathrm{tr}}
+ \frac{1}{2^n}\sum_m\left\|  \Gamma_{\sfF.\enc}^{K} (\ket{m}\bra{m}-\ket{0}\bra{0})\right\|_{\mathrm{tr}}\\
\leq& 2^n \epsilon,
\end{align*}
where the last inequality follows from Lemma~\ref{lemma:offdiagonal} and the assumption of weak IT-security.
Hence $F\left( \rho_{AB}, \sigma_{AB}\right)\geq 1-2^n\epsilon.$
Following steps similar to the proof of the previous theorem, we have
\[
\log|K|\geq 2n+ \log\left(1-\left(\gamma+\sqrt{2^{n}\epsilon}\right)\right).
\]
\hspace*{\fill} $\Box$
\end{proof}

\section{Discussion}\label{sec:discussion}
In this paper we studied quantum encryption with imperfect correctness and imperfect secrecy.
We discussed two notions of information-theoretic security and we showed that the weak security  (Def.~\ref{def:weakITS}) implies the regular security (Def.~\ref{def:strongITS2}) but with a secrecy loss in the dimension of the encrypted message (Theorem~\ref{thm:weaktostrong}).
Two examples of the Ambainis-Smith $\delta$-biased set scheme and the quantum homomorphic encryption scheme for Clifford circuits
were given for applications of Theorem~\ref{thm:weaktostrong}.
We  also showed that an imperfect quantum encryption scheme that encrypts $n$ qubits
can encrypt $2n$ classical bits with the same correctness and secrecy errors (Theorem~\ref{lemma:qtoc}).

The quantum Shannon impossibility results were generalized to the case with imperfect secrecy and imperfect correctness
 for both notions of security.
 For $\epsilon \ll 2^{-n}$, a weak ITS quantum encryption scheme is also ITS by Theorem~\ref{thm:weaktostrong}. 
In this case, our Shannon impossibility results for weak ITS scheme also agrees that roughly $2n$ bits are necessary to encrypt $n$ qubits.
 For regular ITS quantum encryption schemes
 our lower bound (Theorem~\ref{thm:shannon}) is slightly lower than the one by Desrosiers and Dupuis in Eq.~(\ref{eq:SI2})  in the case of $\gamma=0$.
This is because we have to use fidelity and trace distance alternatively in the proof, using the inequalities of Fuchs and van de Graaf~Eq.~(\ref{eq:FT}),
which induces a loss in the secrecy error.

\section*{Acknowledgements}
We are grateful to anonymous referees for their constructive comments on this manuscript.
CYL was was financially supported from the Young Scholar Fellowship Program by Ministry of Science and Technology (MOST) in Taiwan, under
Grant MOST107-2636-E-009-005.
KMC was partially supported by 2016 Academia Sinica Career Development Award under Grant
No. 23-17 and the Ministry of Science and Technology, Taiwan under Grant No. MOST 103-2221-E-001-022-MY3.


\end{document}